\documentclass[aps,prl,twocolumn,superscriptaddress,floatfix,nofootinbib]{revtex4-2}
\usepackage{graphicx}
\usepackage{amsmath,amssymb,bm}
\usepackage{xcolor}
\usepackage[colorlinks=true,linkcolor={blue!45!black},citecolor={blue!45!black},urlcolor={blue!45!black}]{hyperref}
\graphicspath{{figures/}}
\newcommand{\kvac}{\kappa_{\mathrm{vac}}}
\newcommand{\Einf}{E_{\infty}}
\newcommand{\dc}{\delta_{c}}
\newcommand{\ds}{\delta^{*}}
\newcommand{\bS}{\mathbf{S}}

\begin{document}

\title{Domain walls of a spin-chirality ferroelectric as tunable quantum particles}

\author{A.~Buzdakov}
\affiliation{Quantum Materials Theory, Genoa 16162, Italy}
\author{A.~Hingane}
\affiliation{Indian Institute of Science, Bengaluru 560012, India}
\author{S.~Artyukhin}
\affiliation{Quantum Materials Theory, Genoa 16162, Italy}

\begin{abstract}
In frustrated spin chains with competing nearest- and next-nearest-neighbor exchange, the vector spin chirality sets the electric polarization, so a ferroelectric domain wall is a domain wall of the chirality.
The semiclassical picture suggests nonlocal switching dynamics due to topological constraints on the spins, but little is known about such domain walls in a quantum spin chain.  
Using density-matrix renormalization group calculations for a frustrated spin-$\frac12$ XXZ
chain in which a bond alternation $\delta$ opens a spin gap, we show that the wall then becomes a quantum particle.
In a finite chain, it has the energy levels and wave functions of a particle in a box, and it moves under an electric field as a particle under a constant force does. In the infinite chain, its mass grows by more than two
orders of magnitude as $\delta$ decreases and diverges at $\ds\approx0.012$, where the bottom of its band becomes flat;
below $\ds$ the band has two degenerate minima away from its center.
The large mass results from a near cancellation of in-plane and out-of-plane exchange energies.
\end{abstract}

\maketitle

In spiral magnets the electric polarization is set by the vector spin chirality,
$\mathbf{P}\propto\mathbf{e}_{ij}\times(\bS_i\times\bS_j)$, where $\mathbf{e}_{ij}$ connects neighboring
sites~\cite{Katsura2005,Mostovoy2006,Sergienko2006,Tokura2014}. Edge-sharing cuprate chains such as LiCu$_2$O$_2$ and
LiCuVO$_4$ combine a ferromagnetic nearest-neighbor exchange $J_1<0$ with an antiferromagnetic next-nearest-neighbor
exchange $J_2>0$~\cite{Masuda2004,Enderle2005}, and their polarization follows the spin
chirality~\cite{Park2007,Naito2007,Schrettle2008,Seki2008}. The two signs of the chirality $\kappa$ correspond to the two
ferroelectric domains, so a ferroelectric domain wall in these materials is a domain wall of the chirality. In spin-driven
ferroelectrics, domain walls are mobile, and their motion governs the switching of the polarization and contributes
strongly to the dielectric response~\cite{Kagawa2009,Niermann2014,Dragicevic2021}.

\begin{figure*}[t]
\includegraphics[width=\textwidth]{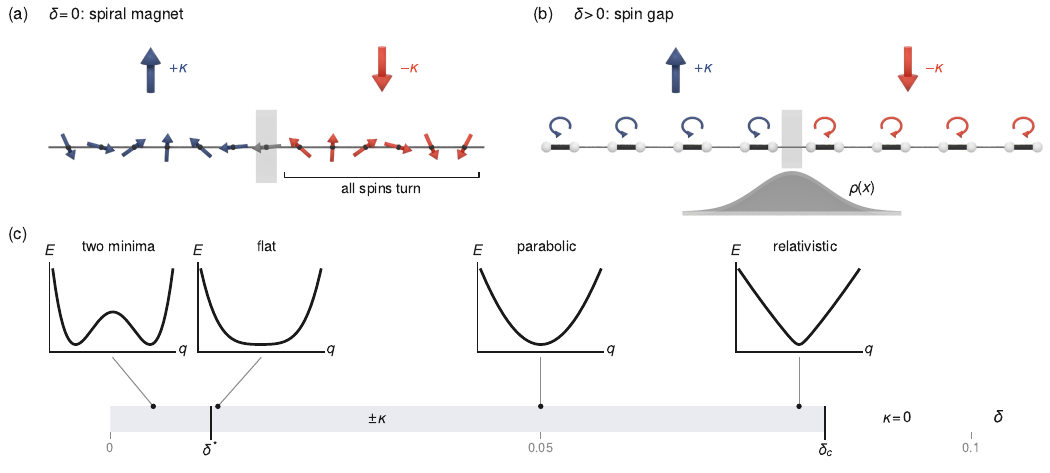}
\caption{(a) Without a spin gap ($\delta=0$) the spins form a spiral whose sense of rotation, the chirality $\kappa$, sets
the polarization (large arrows). The domain wall (gray) is bound to the spiral: moving it turns all spins on one side.
(b) With a spin gap ($\delta>0$; thick lines, stronger ferromagnetic bonds) only the chirality is ordered (curled
arrows), and the wall is a quantum particle with a position distribution $\rho(x)$. (c) Band $E(q)$ of the wall at
$\delta=0.005$, 0.0125, 0.05, and 0.08 (computed at $\chi=256$, each scaled to its own height), placed above the $\delta$
axis with $\ds=0.0117$ and $\dc=0.0830$.}
\label{fig:concept}
\end{figure*}

In a spiral magnet with a strong easy-plane anisotropy, the spins stay in the easy plane across the wall, and the motion of the wall is tied to the phason mode, which rotates all the spins of a domain.
Therefore, moving the wall by one lattice spacing rotates every spin in at least one of the adjacent domains~\cite{Kolezhuk2000,Foggetti2026} [Fig.~\ref{fig:concept}(a)], so such a wall is a macroscopic object whose inertia,
like the D\"oring mass of a magnetic domain wall~\cite{Doring1948}, follows from classical dynamics: in spiral
multiferroics, such walls move as relativistic massive particles with a speed limited by the magnon
velocity~\cite{Maranzana2026}. A single quantum
chain without a spin gap is a critical spin liquid with long-range chiral
order~\cite{Nersesyan1998,Hikihara2001,McCulloch2008,Furukawa2010,Furukawa2012}, which weak interchain coupling turns into
a spiral~\cite{Furukawa2010}. Field theory predicts massive kinks between its two chiral
vacua~\cite{Nersesyan1998,Lecheminant2001,Zarea2004}, but their mass and band have not been computed. The chiral solitons
invoked for the gigahertz dielectric response of LiCuVO$_4$~\cite{Furukawa2008,Grams2022}, and those recently quantized in
chains with a Dzyaloshinskii--Moriya interaction~\cite{Chinellato2026}, are different objects: they leave the chirality
on both of their sides unchanged.

A spin gap removes the spiral, and the chirality can then order even without long-range magnetic order~\cite{Furukawa2012,Ueda2014}. In a single chain, an alternation of the ferromagnetic bonds, $J_1(1\pm\delta)$, opens such a gap in the chiral phase and
produces two vector-chiral dimer phases~\cite{Ueda2014,Ueda2020b}. Two inequivalent ferromagnetic bonds per chain occur,
for example, in Rb$_2$Cu$_2$Mo$_3$O$_{12}$ and LiCuSbO$_4$~\cite{Ueda2020a,Grafe2017}. Rb$_2$Cu$_2$Mo$_3$O$_{12}$ has a
spin gap and was proposed to be ferroelectric without magnetic order~\cite{Reynolds2019,Ueda2020a}, although measurements
on single crystals assign the dielectric anomaly behind this proposal to an impurity phase~\cite{Hayashida2021}. With a gap, the chirality wall is no longer bound
to spiral phasons [Fig.~\ref{fig:concept}(b)] and can be an elementary excitation of the chain.
Domain-wall-like elementary excitations (kinks) are known in
other spin chains. In an Ising-like antiferromagnetic chain, a single wall propagates in a
band~\cite{Villain1975,Nagler1982} and, in a chain with fixed end spins, behaves as a free particle in a
well~\cite{Mikeska1991}. In a biaxial Ising-like ferromagnetic chain, a field along the easy axis is predicted to make a
single wall perform Bloch oscillations~\cite{Kyriakidis1998}, which were observed by neutron scattering for the two
walls of short clusters of reversed spins~\cite{Hansen2022}. In a frustrated, spontaneously dimerized chain, a spinon
is a wall between the two dimer patterns~\cite{Shastry1981}, and a single spinon in an open chain of odd length forms
approximately a particle-in-a-box state~\cite{Sorensen1998}. For ferroelectric walls, quantum motion has so far been
inferred only in organic ferroelectrics: from athermal creep~\cite{Kagawa2016} and from athermal relaxation of spin
solitons~\cite{Imajo2021}, which are the walls of a ferroelectric spin-Peierls state and were argued to form a quantum
liquid~\cite{Imajo2024}. Whether a ferroelectric domain wall in a gapped vector-chiral phase can be described as a quantum particle, and what sets its mass, have not been studied.

Here we show that it can. 
The alternation tunes the mass of the wall over more than two orders of magnitude and makes it diverge at a critical value $\ds$.
We use the density-matrix renormalization group (DMRG)~\cite{White1992,Schollwock2011} for finite segments between infinite vacua~\cite{Phien2012} and a plane-wave ansatz for the wall in the infinite chain~\cite{Haegeman2012,HaegemanOV2013,ZaunerStauber2018,Vanderstraeten2019}. %
The wall has the energy levels and wave functions of a particle in a box and responds to an electric
field as a particle would to a constant force. Its dispersion evolves from that of a light relativistic kink near the Ising transition at $\dc$ to a band with a flat bottom at $\ds\approx0.012$.
Its large mass is the result of a near cancellation between exchange energies of opposite sign. Two minima emerge below $\ds$ [Fig.~\ref{fig:concept}(c)].

\textit{Model and vacua.} We consider the chain
\begin{align}
H&=\sum_j J_1\bigl[1-(-1)^j\delta\bigr]h_{j,j+1}+J_2\sum_j h_{j,j+2},\nonumber\\
h_{ij}&=S^x_iS^x_j+S^y_iS^y_j+\Delta\,S^z_iS^z_j,
\label{eq:H}
\end{align}
with $J_1=-2.5$ for the alternating nearest-neighbor exchange, $J_2=1$ as the unit of energy, and $\Delta=0.8$, for which the uniform chain ($\delta=0$) is in the
vector-chiral phase~\cite{Furukawa2012}. The order parameter is the chirality of the nearest-neighbor bonds,
\begin{equation}
\kappa=\bigl\langle(\bS_j\times\bS_{j+1})^z\bigr\rangle
=\tfrac{i}{2}\bigl\langle S^+_jS^-_{j+1}-S^-_jS^+_{j+1}\bigr\rangle.
\label{eq:kappa}
\end{equation}
For $0<\delta<\dc=0.0830$ the chain has two degenerate gapped vacua with chirality $\pm\kappa$.
The chirality is $\kappa\approx0.17$ at small $\delta$ and vanishes at $\dc$ as $(\dc-\delta)^\beta$ with $\beta=0.122$, close to the
Ising exponent $1/8$~\cite{SM}. For $\delta\le0.02$ the alternation changes $\kappa$ by less than 1\%, so it leaves the
chirality of the uniform chain almost intact while it opens the spin gap.

\begin{figure*}[t]
\includegraphics[width=\textwidth]{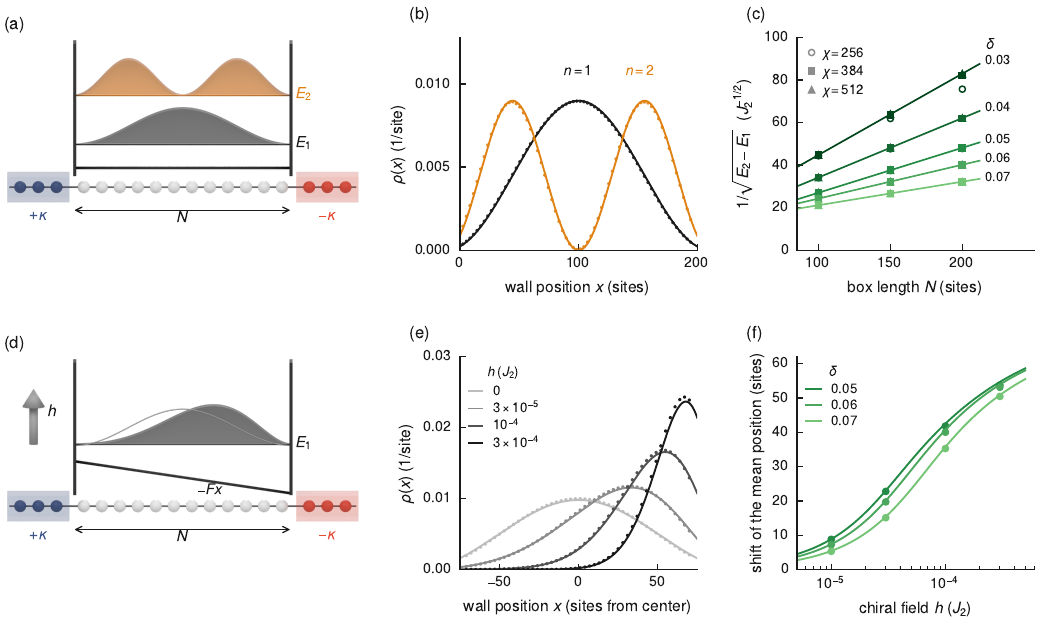}
\caption{(a) $N$ sites between the frozen $+\kappa$ and $-\kappa$ vacua hold one wall, which moves as a particle in a box
with levels $E_1$ and $E_2$ (schematic). (b) Position distribution $\rho(x)$ in the ground ($n=1$) and first excited
($n=2$) states for $\delta=0.04$, $N=200$, $\chi=512$ (points), and $\sin^2(n\pi x/L)$ with $L=N+b$ from the gaps
(lines). (c) $1/\sqrt{E_2-E_1}$ versus $N$ for $\delta=0.03$--0.07 at $\chi=256$ (open), 384, and 512 (filled); lines,
Eq.~(\ref{eq:box}). (d) A chiral field $h$ tilts the floor of the box by $-Fx$ (schematic). (e) $\rho(x)$ for
$\delta=0.06$, $N=150$, $\chi=256$ at several $h$ (points) and the particle in the tilted box (lines). (f) Shift of the
mean position of the wall versus $h$ (points) and the same model (lines). The lines in (e) and (f) use $m$ and $b$ from
the gaps and $F=2h\kvac$, with no fit.}
\label{fig:particle}
\end{figure*}

\textit{Particle in a box.} We use DMRG to optimize $N$ sites between the $+\kappa$
vacuum on the left and the $-\kappa$ vacuum on the right, both kept fixed as infinite boundary conditions, so that the
segment contains exactly one wall [Fig.~\ref{fig:particle}(a)]. A particle of mass $m$ in a box of effective length
$L=N+b$ has the levels
\begin{equation}
E_n=\Einf+\frac{n^2\pi^2}{2mL^2},\qquad n=1,2,\dots,
\label{eq:box}
\end{equation}
with $\hbar=1$ and lengths in lattice sites, so that $1/\sqrt{E_2-E_1}=\sqrt{2m/3\pi^2}\,(N+b)$ grows linearly with $N$.
This is what we find for $\delta=0.03$--0.07 and $N=100$--200 [Fig.~\ref{fig:particle}(c)], and the slope and intercept
give $m$ and $b$. These two numbers then predict where the wall is, with no free parameters. We measure its position
distribution as
\begin{equation}
\rho(x)=-\tfrac12\,\partial_x\!\left[\kappa(x)/\kvac\right]:
\label{eq:cloud}
\end{equation}
for a sharp wall at $X$ the chirality jumps from $+\kvac$ to $-\kvac$ at $X$, so $\rho(x)$ is the probability density of
$X$, broadened by the core of the wall. In the ground and first excited states it agrees with $\sin^2(n\pi x/L)$ to within
0.5\% and 1.1\% of its maximum [Fig.~\ref{fig:particle}(b)], and in the excited state the wall avoids the center of the
box, where the standing wave has its node. The broad chirality profile of the segment is thus the image of a delocalized
particle. The wall itself is narrow: its core has a center of less than four sites, and 2\% of the vacuum chirality is
still missing 13 to 28 sites from the center for $\delta=0.03$--0.07~\cite{SM}.

\textit{Wall in a chiral field.} Next, we apply a force to the wall. Through
the spin-current mechanism an electric field $E_y$ normal to the chain and to the chirality axis couples linearly to the chirality~\cite{Katsura2005,Mostovoy2006}, which we model by a uniform chiral field,
\begin{equation}
H_h=-h\sum_j\kappa_j,
\label{eq:field}
\end{equation}
acting on the $J_1$ bonds of the segment. Moving the wall by one site to the right turns one bond from $-\kvac$ to
$+\kvac$, so the field exerts a constant force $F=2h\kvac$ and tilts the floor of the box [Fig.~\ref{fig:particle}(d)].
The particle of Eq.~(\ref{eq:box}) in the potential $-Fx$, with $m$ and $b$ taken from the gaps, then predicts both the
shape of the position distribution and its displacement. Both agree with DMRG for $h=10^{-5}$ to $3\times10^{-4}$, over
which the field moves the wall by up to 54 sites [Fig.~\ref{fig:particle}(e) and \ref{fig:particle}(f)]. At
$\delta=0.06$ the distributions agree to within 4\% of their maximum, and the displacement found with DMRG is 1\%--2.5\%
smaller than predicted, independently of $\chi$~\cite{SM}; we have not identified the origin of this small deviation.

\begin{figure*}[t]
\includegraphics[width=\textwidth]{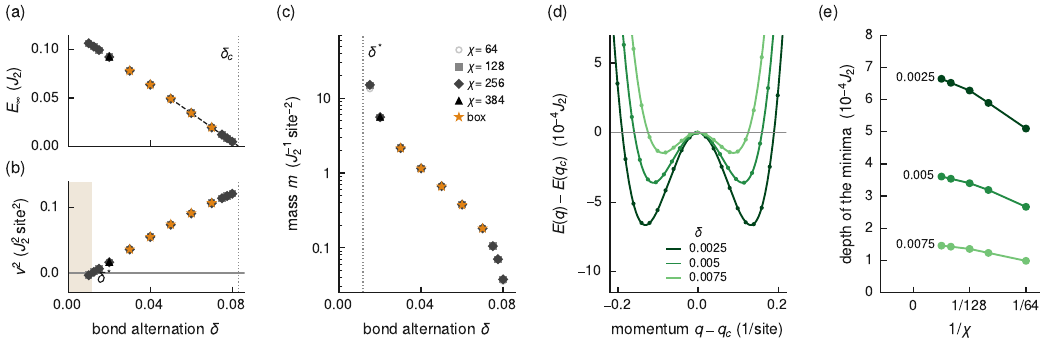}
\caption{(a) Rest energy $\Einf$ and (b) $v^2$ of the band, Eq.~(\ref{eq:band}), at $\chi=64$ to 384 (gray scale) and
from the box (stars). Dashed line in (a), linear fit for $\delta=0.07$--0.08, which vanishes at 0.0831; dotted lines,
$\dc$. In (b) $v^2$ vanishes at $\ds=0.0117$; shaded, two minima. (c) Mass $m=\Einf/v^2$; at $\delta=0.0125$ the mass
still changes by 12\% between $\chi=128$ and 256 and is not shown. (d) Band near its center for $\delta=0.0025$, 0.005,
and 0.0075 at $\chi=256$ (points) and fits (lines). (e) Depth of the two minima versus $1/\chi$.}
\label{fig:band}
\end{figure*}

\textit{Band and mass.} In the infinite chain the particle has a band, which we compute directly with the plane-wave ansatz.
Close to its center $q_c$,
\begin{equation}
E(q)^2=\Einf^2+v^2(q-q_c)^2+O[(q-q_c)^4],\qquad m=\Einf/v^2,
\label{eq:band}
\end{equation}
which we fit at every $\delta$ and $\chi$~\cite{SM}. For $\delta=0.03$--0.07 the band and the box give the same mass to
within 0.9\% and the same $\Einf$ to within $7\times10^{-6}$ [Fig.~\ref{fig:band}]. Standing waves in a finite segment
and plane waves in an infinite chain thus describe the same particle.

The dependence on $\delta$ has two special points [Fig.~\ref{fig:band}(a) and \ref{fig:band}(b)]. Near $\dc$ the
rest energy $\Einf$ vanishes linearly and extrapolates to zero at $\delta=0.0831$, in agreement with $\dc$ from the
chirality and with the Ising exponent $\nu=1$. There the wall is a light relativistic kink, as in the quantum Ising
chain~\cite{Sachdev2011}, and its quantum size $\lambda=v/\Einf$ grows to 76 sites at $\delta=0.08$. In a box shorter than
$\pi\lambda/\sqrt8\approx1.1\lambda$ the confinement energy of one wall exceeds $2\Einf$, the energy of an additional pair
of walls, and the single-particle description fails~\cite{SM}. As $\delta$ decreases, $\Einf$ grows, while $v^2$ falls
almost linearly and vanishes at $\ds=0.0117$. The mass therefore increases from 0.038 at $\delta=0.08$ to 15 at
$\delta=0.015$ (in units of $J_2^{-1}$ per site$^2$) [Fig.~\ref{fig:band}(c)] and diverges at $\ds$, where the bottom of
the band is flat, $E-\Einf\propto(q-q_c)^4$ [Fig.~\ref{fig:concept}(c)].

Below $\ds$ the coefficient $v^2$ is negative, and the band has two minima, at a distance $q^*$ on either side of its
center [Fig.~\ref{fig:band}(d)]. The
minima deepen with increasing $\chi$, with increments that decrease by about a factor of 2 per step
[Fig.~\ref{fig:band}(e)]. At $\chi=256$ their depths are 6.6, 3.6, and $1.5\times10^{-4}$, at $q^*=0.13$, 0.11, and
0.088, for $\delta=0.0025$, 0.005, and 0.0075.
At $\ds$ the minimum of the band thus splits in two, as the minimum of the Landau free energy does when its gradient term changes sign~\cite{Hornreich1975}. Time reversal maps $q_c+q$ onto $q_c-q$ at fixed chirality, so the two minima are exactly degenerate. Near each of them the band is again parabolic, so the mass diverges only at $\ds$. A dispersion that turns quartic and then develops two minima is known for the magnons of the spin-1 bilinear-biquadratic chain, where $v$ also vanishes~\cite{Golinelli1999}, and for the spinons of the frustrated Heisenberg chain, which are kinks between dimer patterns~\cite{Brehmer1998,Lavarelo2014}.

\begin{figure}[!t]
\includegraphics[width=\columnwidth]{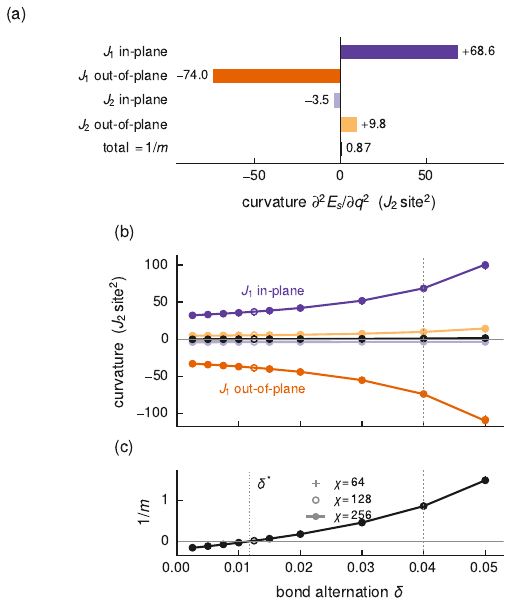}
\caption{(a) Curvatures of the energies of the four types of bonds at the band center, Eq.~(\ref{eq:budget}), and their
sum $1/m$ at $\delta=0.04$, $\chi=256$. (b) The same versus $\delta$ for $\chi=64$, 128, and 256; dashed line,
$\delta=0.04$. (c) The sum $1/m$ on a finer scale; it changes sign at $\ds$.}
\label{fig:budget}
\end{figure}

\textit{Origin of the large mass.} The Hamiltonian is a sum of four types of terms, the in-plane ($S^xS^x+S^yS^y$) and
out-of-plane ($\Delta S^zS^z$) parts of the $J_1$ and $J_2$ bonds. The energy of the wall at momentum $q$ splits exactly
into the corresponding contributions $E_s(q)$, each measured from the vacuum, and so does the inverse mass,
\begin{equation}
\frac1m=\frac{\partial^2E}{\partial q^2}\bigg|_{q_c}=\sum_s\frac{\partial^2E_s}{\partial q^2}\bigg|_{q_c}.
\label{eq:budget}
\end{equation}
Figure~\ref{fig:budget}(a) shows the contributions of the in-plane and out-of-plane $J_1$ and $J_2$ terms to the curvature (inverse mass). The nearest-neighbor exchange $J_1$ gives the largest contributions, but the in-plane and out-of-plane contributions almost compensate each other: at $\delta=0.04$ the in-plane contribution is $+68.6$ and the out-of-plane contribution, from $S^z_iS^z_{i+1}$, is $-74.0$. The next-nearest-neighbor exchange contributes less, and the total, $1/m=0.87$, is about 1\% of either large term. The same cancellation holds for all
$\delta$ shown in Fig.~\ref{fig:budget}(b), where $m^{-1}$ never exceeds 2\% of either large term. The curvature at the band center changes sign at $\ds$ [Fig.~\ref{fig:budget}(c)] and below $\ds$ no longer gives the mass of the lowest states. Since $J_1<0$, the signs of the two large terms mean that as the momentum of the wall grows, its nearest-neighbor bonds gain ferromagnetic $S^z_iS^z_{i+1}$ correlations at the expense of the easy-plane correlations.
This appears to be a quantum counterpart of the D\"oring mass~\cite{Doring1948} of a classical ferromagnetic domain wall, where the spins of a moving wall tilt out of the easy plane, the resulting magnetization plays the role of the wall momentum, and the easy-plane anisotropy that counteracts the tilt sets the inverse mass. There, the wall is heavy when the tilt costs little energy, just as in our case, where the ferromagnetic out-of-plane exchange almost cancels this cost. A conjecture that links $\ds$ to such a tilt is given in the End Matter. %

\textit{Conclusions and outlook.} For $\delta>\ds$ the spin gap thus turns the domain wall of a frustrated chain from a defect bound to the phasons of the adjacent domains into a massive quantum particle, on which an electric field $E_y$ exerts a constant force. Such a wall forms standing waves in a finite sample and has a mass controlled by $\delta$. Because the inverse mass is a small remainder of nearly canceling exchange energies, a bond alternation $\delta$ of a few percent tunes the wall from a light relativistic kink near $\dc$ to a particle with a quartic band bottom and an infinite mass ($m^{-1}=0$) at $\ds$. Below $\ds$ the band has two degenerate minima at $q_c\pm q^*$, where the mass is finite again.

Bloch oscillations of quantum ferromagnetic domain walls were predicted in a magnetic field~\cite{Kyriakidis1998} and observed for walls in clusters of spins in CoCl$_2\cdot$2D$_2$O~\cite{Hansen2022} and for a single kink in a trapped-ion simulator~\cite{De2026}, whereas quantum motion of ferroelectric walls has been inferred from their athermal dynamics in organic ferroelectrics~\cite{Kagawa2016,Imajo2021}.
The wall found here is a ferroelectric domain wall with a band, and an electric field $E_y$ acts on it directly as a constant force. A particle in a band of finite width $W$ cannot move indefinitely under such a force without dissipation, because it can convert at most $W$ of potential energy into kinetic energy. Provided the band of the wall stays separated from other excitations across the Brillouin zone, a weak static electric field should therefore localize the wall in a clean chain, with a Wannier--Stark ladder of levels and Bloch oscillations. Near $\ds$, where the bottom of the band is flat, weak perturbations such as disorder or interchain coupling are expected to control the motion of the wall, and below $\ds$ the two band minima predict standing-wave fringes with period $\pi/q^*$ in the position distribution of a wall in a finite chain, which would test the band independently.

\begin{acknowledgments}
\end{acknowledgments}

\section*{End Matter}

\textit{Units and symmetry.} Energies are in units of $J_2$, lengths in lattice sites, and $\hbar=1$. Since $J_1<0$, the
bond $J_1(1+\delta)$ is the stronger ferromagnetic bond. All calculations conserve the total $S^z$, and the wall has
$S^z=0$.

\textit{Vacua.} The vacua are infinite matrix-product states with a four-site unit cell, obtained with infinite-system DMRG
in TeNPy~\cite{Schollwock2011,Hauschild2018} for $\chi=64$ to 1024. Each vacuum is found by iterating first in a weak
chiral field $h=\pm0.05$, which selects the sign of $\kappa$, and then at $h=0$. The energies per site of the two vacua
agree to $10^{-9}$ and their chiralities are opposite to within $10^{-6}$. $\dc$ and $\beta$ follow from a fit of
$\kappa=A(\dc-\delta)^\beta$ for $\delta=0.0725$ to 0.0825.

\textit{Segments.} $N$ sites are optimized with DMRG between the two vacua ($\chi=256$), which enter as fixed boundary
environments~\cite{Phien2012}; the segment has $\chi=256$, 384, and 512, and 768 for $N=100$ at $\delta=0.03$ and 0.04. A
run stops when the energy changes by less than $10^{-10}$ and the largest bond entropy by less than $10^{-6}$ over ten
sweeps. The first excited state is the lowest state orthogonal to the ground state. $\rho(x)$ is evaluated on the $J_1$
bonds as a central difference over four sites, which removes the modulation caused by the alternation, and $m$ and $b$
follow from a least-squares fit of $1/\sqrt{E_2-E_1}$ at $N=100$, 150, and 200. In the field model, the Schr\"odinger
equation of the particle in the box with potential $-Fx$ is solved numerically, $\kvac$ is the mean chirality of the
vacuum on the $J_1$ bonds, and the displacement is the change of the mean of $\rho(x)$ over $|x|\le N/2-3$.

\textit{Band.} The wall in the infinite chain is a plane-wave superposition of a local tensor placed between the $+\kappa$
and $-\kappa$ vacua~\cite{Haegeman2012,HaegemanOV2013,ZaunerStauber2018,Vanderstraeten2019}. The momentum of a single wall is defined only up to a constant~\cite{Vanderstraeten2019}. Here momenta are measured from
$k_0=\arg\Lambda/L_{\mathrm{cell}}$, where $\Lambda$ is the leading eigenvalue of the mixed transfer matrix of the two
vacua, which fixes the relative phase of the independently obtained vacua. The implementation reproduces the exact
domain-wall dispersion of the transverse-field Ising chain to $10^{-14}$, and extending the local tensor to four sites
changes the depth of the two minima by 1\%. We fit $E^2=\Einf^2+v^2(q-q_c)^2+\gamma(q-q_c)^4+\zeta(q-q_c)^6$ for
$|q|\le0.16$ with a free center $q_c$, at $\chi=64$, 128, and 256 for all $\delta$ and also at $\chi=384$ for
$\delta=0.02$. $\ds$ is the zero of $v^2$, interpolated linearly between $\delta=0.01$ and 0.015. For $\delta<\ds$ the
band is computed on dense momentum grids and fitted near its center with an even polynomial of sixth order, which gives
$q^*$ and the depth of the minima.

\textit{Bond decomposition.} The four parts of $H$ are Hermitian, so the plane-wave energy of the wall splits exactly into
four expectation values $E_s(q)$. Their curvatures at $q_c$ follow from central differences with step 0.02 and Richardson
extrapolation, and their sum agrees with the curvature of the band to $10^{-13}$. The in-plane and out-of-plane terms of
the $J_1$ bonds change with $\chi$ by at most 0.7\% for $\delta\le0.04$ and 1.3\% at $\delta=0.05$. The separate totals of
the $J_1$ and $J_2$ bonds still change by up to 30\%--40\% between $\chi=128$ and 256 for $\delta<0.02$, although their
signs and their sum do not; we therefore compare these totals only at $\delta=0.04$.

\textit{Symmetry and the two minima.}
Time reversal leaves the chirality unchanged and reverses $S^z$ and $q-q_c$. The band is therefore even in $q-q_c$, and near $\ds$ it has the form $E-\Einf=c_2(q-q_c)^2+c_4(q-q_c)^4$ with $c_2=1/(2m)$ and $c_4>0$. For $c_2<0$ the minima are at $q^{*2}=-c_2/(2c_4)$, their depth is $c_2^2/(4c_4)$, and the curvature at each of them is $-4c_2$, so the mass there is $1/(4|c_2|)$. With $c_2$ linear in $\delta-\ds$, $q^{*2}$ grows linearly with $\ds-\delta$, as the three values of $q^*$ quoted above do to within their rounding.

\textit{Conjecture on the origin of $\ds$.} We conjecture that the momentum of the wall is carried by a tilt of the spins out of the easy plane, as for the D\"oring wall. Time reversal allows a wall with momentum $q_c+q$ to carry a distribution of $S^z$ that is odd in $q$, and since the wall has $S^z=0$ this distribution sums to zero. Without a spin gap the wall is bound to the phasons of the adjacent domains [Fig.~\ref{fig:concept}(a)]. With a gap only this local tilt would remain, and it would set the inertia of the wall. A larger gap would make the tilt more costly and the wall lighter, as we find for increasing $\delta$. In this picture $\ds$ is the alternation below which the tilt lowers the energy of the wall, and the two minima are walls tilted in opposite senses. The picture makes two predictions that we have not tested. A time-reversal-odd quantity of the wall, such as its scalar spin chirality, is odd in $q-q_c$ and nonzero at the two minima, and $\ds$ grows as $\Delta$ approaches 1.

\textit{Absence of spontaneous dimerization.} The energy gain from the alternation grows as $\delta^2$, with
$[e(0)-e(\delta)]/\delta^2\to0.90$ as $\delta\to0$, both from the total energy and from the difference between the
energies of strong and weak bonds, which agree to 0.2\% ($\chi$ up to 1024). A gain proportional to $\delta^2$ does not
lead to a spontaneous dimerization at weak coupling to the lattice, so the alternation has to be present in the crystal
structure.

\end{document}


\title{Supplemental Material for ``Domain walls of a spin-chirality ferroelectric as tunable quantum particles''}
\author{A.~Buzdakov}
\affiliation{Quantum Materials Theory, Genoa 16162, Italy}
\author{A.~Hingane}
\affiliation{Indian Institute of Science, Bengaluru 560012, India}
\author{S.~Artyukhin}
\affiliation{Quantum Materials Theory, Genoa 16162, Italy}
\maketitle

\section{Vacua}
\begin{figure}[h]
\includegraphics[width=0.5\textwidth]{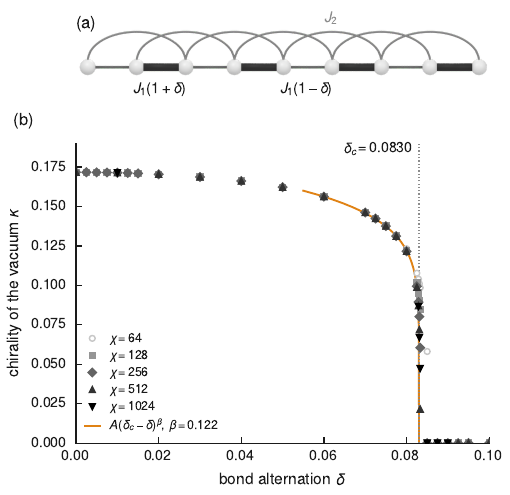}
\caption{(a) The chain of Eq.~(1) of the Letter: nearest-neighbor bonds alternate between $J_1(1+\delta)$ (thick) and
$J_1(1-\delta)$ (thin); arcs, next-nearest-neighbor bonds $J_2$. (b) Chirality $\kappa$ of the vacuum versus $\delta$
for $\chi=64$ to 1024 (gray scale). Orange line, fit of $A(\dc-\delta)^\beta$ at $\chi=512$; dotted line, $\dc$.}
\label{fig:vac}
\end{figure}

The chain and its parameters are given in Eq.~(1) of the Letter and drawn in Fig.~\ref{fig:vac}(a). The two vacua are
infinite matrix-product states obtained with infinite-system DMRG~\cite{Schollwock2011,Hauschild2018}; the chirality
$\kappa(\delta)$ is shown in Fig.~\ref{fig:vac}(b) for $\chi=64$ to 1024. The transition point and the exponent follow from
a fit of $\kappa=A(\dc-\delta)^\beta$ for $\delta=0.0725$ to 0.0825. With increasing $\chi$ we obtain $\dc=0.0836$,
0.0832, 0.08304, and 0.08303 for $\chi=64$, 128, 256, and 512, and $\beta=0.122$ at $\chi=512$. At small $\delta$ the
chirality depends only weakly on the alternation, $\kappa=0.1717$, 0.1705, and 0.1662 for $\delta=0$, 0.02, and 0.04 at
$\chi=512$.

\section{Response to the chiral field}
\begin{figure}[h]
\includegraphics[width=\textwidth]{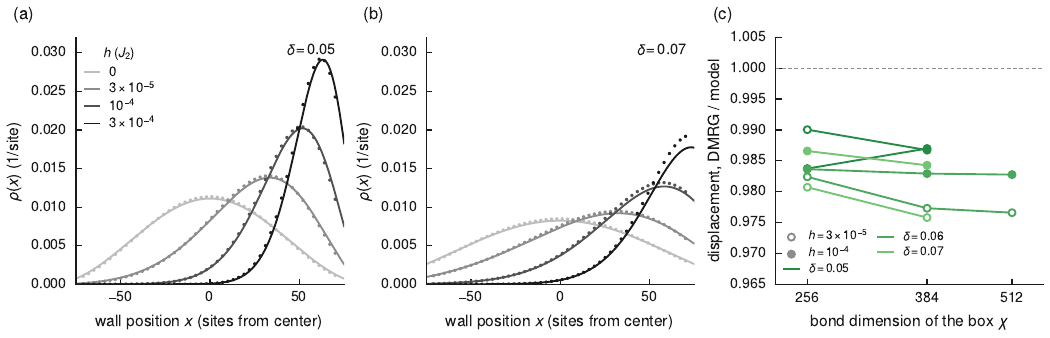}
\caption{(a), (b) Position distribution $\rho(x)$ of the wall at $\delta=0.05$ and 0.07, $N=150$, $\chi=256$ for $h=0$ to
$3\times10^{-4}$ (points) and the particle model with $m$ and $b$ from the gaps (lines, no fit); the case
$\delta=0.06$ is Fig.~2(e) of the Letter. (c) Ratio of the DMRG displacement to the model prediction versus the bond
dimension of the box for $\delta=0.05$, 0.06, and 0.07 (dark to light green); open symbols, $h=3\times10^{-5}$; filled
symbols, $h=10^{-4}$. The vacua have $\chi=256$.}
\label{fig:field}
\end{figure}

Figures~2(e) and 2(f) of the Letter compare the wall in a chiral field with the particle model at $\delta=0.06$ and for
the displacement at $\delta=0.05$--0.07. The model uses the mass $m$ and the box length $L=N+b$ obtained from the gaps at
the same bond dimension, and the force $F=2h\kvac$; nothing is adjusted to the field calculations.
Figures~\ref{fig:field}(a) and \ref{fig:field}(b) show the position distributions at $\delta=0.05$ and 0.07. Over
$h=0$ to $3\times10^{-4}$ the model reproduces them to within 2\%--4\% of their maximum, except at $\delta=0.07$ and
$h=3\times10^{-4}$, where the measured distribution exceeds the model by up to 9\% within five sites of the wall of the
box. Figure~\ref{fig:field}(c) shows the ratio of the displacement found with DMRG to the model prediction as a function
of the bond dimension of the box. The ratio lies between 0.976 and 0.990 and changes by at most 0.5\% per step in $\chi$,
so the deviation of 1\%--2.5\% is not a convergence effect. Its origin is not known.

\section{Length scales of the wall}
\begin{figure}[h]
\includegraphics[width=\textwidth]{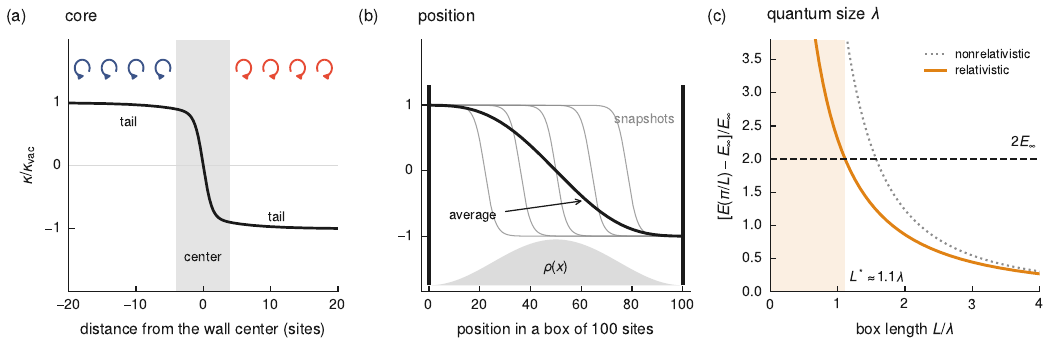}
\caption{Length scales of the wall (schematic). (a) Core: sharp center and slow tails of the chirality profile; curled
arrows, sense of the twist. (b) Position: snapshots of a sharp wall (thin lines), their average (thick line), and the
distribution $\rho(x)$ of the center (shaded). (c) Quantum size: energy needed to hold one wall in a box of length $L$,
from the relativistic band at $q=\pi/L$ (solid line) and for a nonrelativistic particle (dotted line); dashed line,
$2\Einf$; shaded, $L<L^*$.}
\label{fig:lsch}
\end{figure}

Three lengths characterize the wall [Fig.~\ref{fig:lsch}].

\begin{figure}[h]
\includegraphics[width=0.55\textwidth]{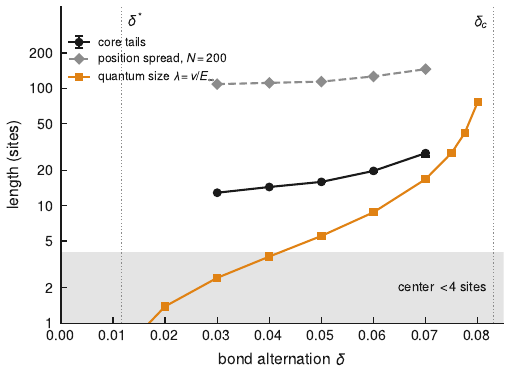}
\caption{Length scales versus $\delta$, with the colors of Fig.~\ref{fig:lsch}. Core: tail length (symbols, largest
segment and $\chi$; bars, spread over $\chi=256$ to 512 and $N=150$ to 300). Position: width of $\rho(x)$ for $N=200$.
Quantum size: $\lambda=v/\Einf$ from the band. Shaded, the range below four sites, in which the center of the core lies
at all $\delta$.}
\label{fig:len}
\end{figure}

\textit{Core.} The chirality profile averaged over the ground state of a segment is the core profile convolved with the
position distribution of the wall. We remove the position distribution exactly by summing the chirality correlations
$\langle\kappa_i\kappa_{i+r}\rangle/\kvac^2$ over all pairs at fixed distance $r$: for any position distribution that
stays away from the ends of the segment, this sum differs from its value for a sharp wall only by a function of $r$ set by
the core profile. Vacuum fluctuations of $\kappa$ are divided out with a segment that contains no wall. The resulting
core has a center narrower than four sites, which the method does not resolve, and tails in which 2\% of the vacuum
chirality is still missing 13, 14, 16, 20, and 28 sites from the center for $\delta=0.03$, 0.04, 0.05, 0.06, and 0.07
[Fig.~\ref{fig:len}]. For $\delta\ge0.075$ this length still grows with the size of the segment and is not shown.

\textit{Position distribution.} Its width is $(N+b)/2$ for the ground state, between 109 and 146 sites for $N=200$, and is
set by the segment.

\textit{Quantum size.} The relativistic band $E=\sqrt{\Einf^2+v^2q^2}$ defines $\lambda=v/\Einf$. Holding one wall in a
box of length $L$ costs
\begin{equation}
E(\pi/L)-\Einf=\Einf\left[\sqrt{1+(\pi\lambda/L)^2}-1\right],
\end{equation}
which exceeds $2\Einf$, the energy of an additional pair of walls, for $L<L^*=\pi\lambda/\sqrt8\approx1.1\lambda$. The
quantum size decreases from 76 sites at $\delta=0.08$ to 17 at 0.07 and 5.5 at 0.05, falls below one site for
$\delta\le0.015$, and vanishes at $\ds$.

\section{Convergence}
Table~\ref{tab:conv} lists the change of each quantity at the last step of its ladder of bond dimensions.

\begin{table}[h]
\caption{Convergence in the bond dimension $\chi$.}
\label{tab:conv}
\begin{ruledtabular}
\begin{tabular}{lll}
Quantity & $\chi$ ladder & Last step \\
\hline
$\dc$ (vacuum) & 64, 128, 256, 512 & $0.08304\to0.08303$ \\
Gap $E_2-E_1$, $N=100$, $\delta=0.03$ & 256, 384, 512, 768 & $-0.06\%$ \\
Mass from the box & 256, 384, 512 & $+0.1\%$ ($\delta=0.07$) to $+4.3\%$ ($\delta=0.03$) \\
Field displacement, DMRG/model & 256, 384, 512 & $\le0.5\%$ \\
Mass from the band, $\delta\ge0.03$ & 64, 128, 256 & $\le0.2\%$ \\
Mass from the band, $\delta=0.02$ & 64, 128, 256, 384 & $+0.1\%$ \\
Mass from the band, $\delta=0.015$ & 64, 128, 256 & $+2.6\%$ \\
$\ds$ & 64, 128, 256 & $0.0116\to0.0117$ \\
Depth of the minima, $\delta=0.0025$ & 64, 96, 128, 192, 256 & $+1.9\%$ \\
In-plane and out-of-plane $J_1$ terms & 64, 128, 256 & $\le0.7\%$ ($\delta\le0.04$), $1.3\%$ ($\delta=0.05$) \\
\end{tabular}
\end{ruledtabular}
\end{table}